# Lick Galaxy Correlation Function Revised

Manolis Plionis[1,2] and Stefano Borgani[1,3]
[1] *Scuola Internazionale Superiore di Studi Avanzati, SISSA, via Beirut 2-4, I-34013 Trieste, Italy*
[2] *International Center for Theoretical Physics, ICTP, Strada Costiera, I-34014 Trieste, Italy*
[3] *INFN Sezione di Perugia, c/o Dipartimento di Fisica dell'Università, via A. Pascoli, I-06100 Perugia, Italy*

**ABSTRACT**
We re-estimate the angular 2-point galaxy correlation function from the Lick galaxy catalogue. We argue that the large-scale gradients observed in the Lick catalogue are dominated by real clustering and therefore they should not be subtracted prior to the estimation of $w(\theta)$. We find that if no such correction is introduced the galaxy correlations are perfectly consistent with the those found in the APM survey. Thus, the long standing discrepancy between the Lick and APM angular correlations is lifted.

**Key words:** galaxies: clustering – galaxies: formation – large–scale structure of Universe

## 1 INTRODUCTION

The Lick galaxy counts (Shane & Wirtanen 1967; Seldner et al. 1977, SSGP hereafter) was the first large wide-angle catalogue to give rise to a lot of statistical studies of galaxy clustering (cf. Limber 1954; Peebles & Hauser 1974; Groth & Peebles 1977, GP77 hereafter). The latter authors calculated, among other things, the two-point angular correlation function of the Lick counts and derived the well known power law form

$$w(\theta) = A\theta^{1-\gamma} \qquad (1)$$

for angular separations $\theta \lesssim 2.5°$, with

$$A \approx 0.0684 \quad \text{and} \quad \gamma \approx 1.741. \qquad (2)$$

Probably the most impressive and much debated feature of $w(\theta)$ is its break at 2.5° corresponding to a spatial scale, at the characteristic depth of the Lick catalogue, of $\sim 10\ h^{-1}$ Mpc. Geller, de Lapparent & Kurtz (1984) and de Lapparent, Kurtz & Geller (1986) argued for an artificial origin of the break, relating it to systematic effects due to plate-to-plate variations of the limiting magnitude and to observer dependent effects. Groth & Peebles (1986a,b, GP86, hereafter) reanalysed the Lick counts, in view of this criticism and concluded that the shape of the 2-point correlation function is not significantly affected by the effects discussed by de Lapparent et al. (1986). Nevertheless, due to this interesting debate, many issues related to plate matching procedures and to observer–dependent effects came to light, which help guide new attempts to construct wide-angle catalogues of extragalactic objects.

With the advent of new automated galaxy surveys (APM, COSMOS), the existence of the break in $w(\theta)$ was tested and has been confirmed. Earlier, Hewett (1982) and Stevenson et al. (1985) had presented this evidence although at that time the scale on which it occurs was in dispute. In the APM galaxy catalogue, the break in the two-point angular correlation function occurs at $\sim 1.5°$ which corresponds to $\sim 2.5° - 3°$ at the depth of the shallower Lick catalogue (Maddox et al. 1990). Furthermore Collins et al. (1989) and (1992) using the COSMOS galaxy survey in the South Galactic hemisphere find consistent results supporting the reality of the break in $w(\theta)$.

Although the issue of the reality of the break in $w(\theta)$ has been clarified, an inconsistency between the Lick and APM correlation functions has emerged. This is the behaviour of $w(\theta)$ at large angular scales, beyond the break. The APM catalogue shows a smoother break which implies more large-scale power. This issue is of great importance since it places severe constrains in models of galaxy formation. In particular the quite popular standard CDM model successfully predicts the Lick map 2-point correlation function, but cannot accommodate the excess power seen in the APM $w(\theta)$ at large $\theta$'s (Bond & Couchman 1988; Moscardini et al. 1993; Baugh & Efstathiou 1993).

Since the volumes probed by both catalogues are large enough to be considered fair samples of the Universe ($V \sim 10^7 - 10^8\ h^{-3}$ Mpc$^3$) the above discrepancy should probably be attributed to errors related to the reduction-estimation and/or plate matching procedure. Nichol & Collins (1993) investigated the effects of galactic extinction and plate–to–plate matching errors on the COSMOS survey. They concluded, however, that the estimated $w(\theta)$ is only marginally affected by such effects, a result pointing in favour of a reliable large-scale power detection. However, Fong, Hale-Sutton & Shanks (1992), discussing the effects of possible unaccounted photometric zero-point errors on the APM $w(\theta)$ showed that removing the effects of such errors re-moves the excess power, at large angular scales, seen in the APM results. In this paper we attack this issue from the opposite side. We argue that one of the procedures used by GP77 to estimate the $w(\theta)$ (ie., large-scale density gradients subtraction) eliminates real clustering, which then results in the sharp break of the correlation function. If all other corrections are included (GP77, GP86) the resulting Lick correlation function is identical to that of the APM survey.

## 2 ESTIMATING THE TWO-POINT CORRELATION FUNCTION

The Lick galaxy counts are contained in 1246 plates covering 8.8 steradians of sky, north of declination $-23°$. Each plate covers an area of $6° \times 6°$ and are spaced along lines of constant declination such that they overlap by at least $1°$ in declination and right ascension. We use the 'free-of-overlap' SSGP corrected counts (Plionis 1988). Note that GP77 and GP86 used the inner $5° \times 5°$ of each plate not taking into account the stretching of the plates with varying declination. Thus they used double counted cells ($\sim 8\%$ of total) while we have not. The 2-point galaxy correlation function has been evaluated using the estimator

$$w_{est}(\theta) = \frac{\langle n_i n_j \rangle_\theta}{\langle \frac{1}{2}(n_i + n_j) \rangle^2_\theta} - 1 \,, \qquad (3)$$

where $\theta - \delta\theta/2 \leq \theta < \theta + \delta\theta/2$ and $n_i$ is the galaxy count of the $i^{th}$ cell. Using the above *local* normalization (Peebles & Hauser 1974) has the advantage of making some correction for inhomogeneities arising from patchy obscuration and from intra-plate gradients (cf. Sharp 1979).

We have estimated $w(\theta)$ using two different cell sizes at two different separation ranges:

- ♠ The original SSGP $10'$ cells with $i$ and $j$ cells from the same plate (intra-plate estimate) for angular separations $0.2° \leq \theta \leq 3.6°$ in linear bins of amplitude $\delta\theta = 0.2°$.
- ♠ $60'$ cells in the separation range $1.6° \leq \theta \leq 52°$. For $\theta \lesssim 6°$ we have used linear bins with $\delta\theta = 0.625°$, while for $\theta > 6°$ we have used logarithmic bins with $\delta(\log \theta) \sim 0.06$. We have estimated the intra-plate $w(\theta)$ for separations $\theta \lesssim 4.8$ and the inter-plate $w(\theta)$ for the whole range of angular separations.

Using the notation of GP86, the correction factor error model is

$$C_i^e = C_i + C_i s_i + \epsilon(C_i - C_\circ) \,, \qquad (4)$$

where $C_i^e$ and $C_i$ are the estimated and real plate correction factors, $\epsilon$ is the amplitude of systematic errors in plate correction factors and $s_i$ is the random plate correction factor error (with $\beta_{ij} = \langle s_i s_j \rangle$ their covariance).

According to this model the intrinsic correlation function $w^*$, is:

$$w^* = (w_{est} - A)/B \,, \qquad (5)$$

with

$$A = \beta_{ij}^2 \,; \qquad (6)$$

$$B = \langle C_i^{\gamma/3-1} C_j \rangle (1 + \beta_{ij}^2) \,. \qquad (7)$$

where we have dropped the $\epsilon$ dependance, since GP86 found $\epsilon = 0 \pm 0.15$, while $\langle C_i^{\gamma/3-1} C_j \rangle \sim 1$ (with a very weak dependance on $\gamma$). For our intra-plate estimates of $w(\theta)$ we use $\beta_{ii} = 0.006$ (as found by GP86) and not the original 0.004 value of GP77 while for inter-plate estimates we use $\beta_{ij} = 0$ (for neighbouring plates having a $\sim 5°$ overlap, GP77 used -0.001, a value that would enhance $w(\theta)$ at the corresponding scales, while GP86 found a $b_{ij}$ value between -0.0015 and 0.001).

The only correction used by GP77, which we have neglected, is the subtraction of the large-scale galaxy density gradients from the map, prior to the estimation of $w(\theta)$. The reasons are the following:

- ♠ Plionis (1988) found that the large-scale number count anisotropies, seen in the Lick map, add up to a robust dipole pointing within $\sim 30°$ of the CMB dipole direction. In fact, this result was the first indication that there is a contribution to the local peculiar motion from scales $\gg 50\ h^{-1}$ Mpc, something which is verified by a number of independent studies (cf. Rowan-Robinson et al. 1990; Scaramella et al. 1991; Plionis & Valdarnini 1991; Plionis, Coles & Catelan 1992; Hudson 1993).
- ♠ Brown & Groth (1989) using elaborate simulations of the Lick map and following the same plate-matching procedure as in SSGP found that the SSGP procedure could introduce large-scale gradients, but by a factor of more than 2 times less than what observed in the Lick map. They concluded that '...*the observed large-scale gradients (amplitude* $\sim$ *20%) in the Lick counts must result from intrinsic gradients in the galaxy distribution...*'. Also '...*one might well expect that a significant fraction of the observed large-scale gradients are due to large-scale gradients in the galaxy distribution...*'

Following GP77, we evaluated the 2-point Lick galaxy correlation function by dividing the whole sample into four zones, corresponding to four galactic latitude intervals: (A) $40° \leq b \leq 55°$; (B) $b > 55°$; (C) $-55° \leq b \leq -40°$; (D) $b < -55°$. We present our estimates of the Lick correlation function in Figure 1. The open symbols denote the corrected intra-plate estimates of the correlation function, based on the $10'$ cells, while the solid circles denote the inter-plate estimates (based on the $60'$ cells). These results are obtained by averaging between the four zones while the errorbars are the corresponding $1\sigma$ scatter. We also plot $w(\theta)$ for $|b| \geq 50°$ (solid line), to test the possible effects of galactic absorption, and it is evident that these estimates are consistent with our main ones.

Note, that the intra- and inter-plate estimates are in close agreement in the overlapping range of angular separations and comparing our $60'$-cell based estimates of the intra- and inter-plate $w(\theta)$ (in the range $1.6° \lesssim \theta \lesssim 4.8°$), we find a mean offset $\langle \delta w \rangle \lesssim 0.004$ (see insert of Figure 1) and a mean fluctuation $\langle \delta w/w \rangle \sim 0.1$. The value of $\langle \delta w \rangle$ corresponds to an rms number count variation among plates of 0.06 which in turn corresponds, assuming $\log N \propto 0.6 m$, to a magnitude offset among plates, $\Delta m \sim 0.04$, which is of the same order as that found in the COSMOS survey (Collins et al. 1992) and $\sim 40\%$ larger than that found in the APM survey (Maddox et al. 1990). The difference between the intra- and inter-plate estimates of $w(\theta)$ is probably the result of the plate-to-plate variation of the limiting apparent

This value of $\gamma$ is consistent with the original GP77 one (see eq.2) while the amplitude $A$ is $\sim 9\%$ higher than the corresponding GP77 one, most probably due to the fact that eliminating the long-wavelength component of the correlations affects the amplitude of $w(\theta)$ even on scales $\sim 1.5° - 2.5°$.

## 3 CONCLUSIONS

We re-analyzed the Lick galaxy angular correlation function with the aim of explaining the reason for the discrepancy found between previous determinations (GP77, GP86) and more recent estimates based on the APM and COSMOS automated surveys (Maddox et al. 1990; Collins et al. 1992). We find that, after including all the corrections in the Lick $w(\theta)$ estimate, as in GP77 and GP86, except that for large-scale gradients, a perfect consistency exists between Lick and APM results, even at scales larger than $2.5°$, at which the $w(\theta)$ power-law shape breaks. Therefore, under the valid, according to our previous discussion, assumption that the major part of the observed large-scale galaxy-density gradients are intrinsic to the galaxy distribution, the long standing discrepancy between the APM and Lick angular correlation functions, at angular scales $\theta \gtrsim 2.5°$ is lifted. Inverting this argument we can also conclude that the consistency of the APM and Lick angular correlation functions favours the physical origin of the large-scale gradients observed in the Lick galaxy counts, as advocated by Plionis (1988) and Brown & Groth (1989), and supports the reliability of clustering analyses based on the Lick sample.

### Acknowledgments

We thank Steve Maddox for supplying the APM galaxy correlation results.

**Figure 1.** The Lick 2-point correlation function, compared with the APM result. Open and filled circles refer to the intra- and inter-plate estimates, respectively. Results are the average between the four subsamples, corresponding to four different galactic latitude intervals (see text). For clarity reasons we plot only a few errorbars (1$\sigma$ scatter between these subsamples). The solid line is the $w(\theta)$ estimate for $|b| \geq 50°$ and the dashed line is the best-fit to the GP77 analysis. Crosses are the APM correlation data, scaled to the Lick depth (Maddox et al. 1990). The insert shows the difference of the $60'$-cell based correlation estimates between inter- and intra-plate analysis, in the angular range of overlap. The errorbars correspond to different values of $b_{ij}$ (in the range [-0.001, 0.001]).

magnitude which introduce correlations between nearby intrinsically faint galaxies, revealed in 'deep' plates, and the bright galaxies at similar distances seen in a neighbouring 'shallow' plate. GP77 discussed this problem and estimated it to have an amplitude of $0.15 w(\theta)$ which is quite close to the value we find. In accordance with GP77 we do not correct for this effect, because its amplitude is uncertain and model dependent [note however that if we correct the inter-plate estimates by $w^*_{inter} = (1-f) w_{inter}$, using $f = 0.1$, we find qualitatively the same results as those before applying this correction].

In Figure 1 we plot, as crosses, the APM correlation function scaled at the depth of the Lick counts (Maddox et al. 1990). As can be seen the two correlation function are perfectly consistent with each other, even at scales larger than $\sim 2.5°$, where the power-law shape of $w(\theta)$ breaks down. This represents our main result, which resolves the long-standing discrepancy between APM and Lick galaxy correlation functions.

We have also fitted the power law of eq.(1) to our results in the range $0° \leq \theta \leq 2.5°$, using a $\chi^2$ minimization in which each value of $w(\theta)$ is weighted by $1/\sigma$, and we obtain:

$$\gamma = 1.753 \pm 0.004 \quad \text{and} \quad A = 0.0746 \pm 0.0004$$